\documentclass{jfm}

\usepackage{graphicx}
\usepackage{upmath}

\ifCUPmtlplainloaded
\else
  \def\upi{\pi} 
\fi
\title[Periodic Steady Vortices in a Stagnation Point Flow II]{Periodic Steady Vortices in a \\ Stagnation Point Flow II}
\author[O. S. Kerr]{ O\ls L\ls I\ls V\ls E\ls R\ns S.\ns K\ls E\ls R\ls R}
\affiliation{Department of Mathematics, City University\footnote{Now City St George’s, University of London. Contact {\tt o.s.kerr@city.ac.uk}.},\\[\affilskip] Northampton Square, London EC1V 0HB, UK}
\date{1997}

\begin{document}

\maketitle

\begin{abstract}
Steady-state perturbations to a stagnation point flow of the form ${\bf U}=(0,A'y,-A'z)$ are known which consist of a periodic array of counter-rotating vortices whose axes are parallel to the $y$-axis and which lie in the plane $z=0$. A new understanding of how these vortices depend on the supply of incoming vorticity from afar has lead to the discovery of new families of steady-state periodic vortices that can exist in a stagnation point flow. These new flows have a greater variety of structures than those previously known.

An understanding of the linkage between the vortices and the weak inflow of vorticity can have important implications for situations where such vortices are observed.
\end{abstract}

\section{Introduction}

A stagnation point flow of the form ${\bf U}=(0,A'y,-A'z)$ has been shown
to be unstable to three-dimensional disturbances (Aryshev, Golovin \&
Ershin, 1981, Lagnado, Phan-Thien \& Leal, 1984, Craik \& Criminale,
1986). In Kerr \& Dold (1994), hereafter referred to as KD, a family of
steady-state nonlinear perturbations to this flow were found that
satisfied the Navier-Stokes equation for an incompressible fluid.
These consist of a periodic row of counter-rotating vortices whose
cores lie on the plane $z=0$, with axes parallel to the $y$-axis, and
which decay as $z\rightarrow\pm\infty$.  These solutions have the
feature that the perturbations to the background stagnation point flow
are uniform in the $y$-direction, and that there is no component of the
perturbation velocity in this direction. In this paper we will look at
some of the features of these steady solutions in more detail, and show
how there is a far more extensive family of steady state perturbations
to this stagnation point flow than those presented in KD.  These
further solutions follow from an improved understanding of the
important r\^ole of vorticity in the formation of these vortices which
was not fully appreciated at the time of writing of KD.

For a more extensive review of the relevant literature see KD.

In section 2 we will look at the way the vorticity decays as
$z\rightarrow\pm\infty$, showing how there must be an algebraically
decaying tail. In section 3 we will present some further families of
steady vortices which will give an indication of the many further
possibilities for solutions.  In section 4 we will look at the
relationship between the algebraically decaying tail of vorticity and
the amplitude, and in section 5 we will discuss  the results, including
implications for growing periodic vortices in a stagnation point flow.

\section{Governing equations}

Any perturbation to a stagnation point flow of the form ${\bf
U}=(0,A'y,-A'z)$ that initially has no $y$-dependency will always have no
$y$-dependency. If such a perturbation initially has no velocity
component in the $y$-direction then it will remain like this.  The
perturbation velocity will then be a 2-dimensional incompressible
motion parallel to the $x$-$z$ plane. The non-dimensional governing
equations for such a perturbation to a stagnation point flow of the
above form derived in KD are
\begin{equation}-\frac{\partial \psi}{\partial z}\frac{\partial
\omega}{\partial x}+\frac{\partial \psi}{\partial x}\frac{\partial
\omega}{\partial z}-\lambda z\frac{\partial \omega}{\partial
z}-\lambda\omega=\left(\frac{\partial^2}{\partial
x^2}+\frac{\partial^2}{\partial z^2}\right)\omega,\label{OmEqn}\end{equation}
where the stream function for the perturbation velocity is $\psi(x,z)$
such that the perturbation velocity is given by
\begin{equation}{\bf u}=(u,0,w)=\left(-\frac{\partial
\psi}{\partial z},0,\frac{\partial \psi}{\partial
x}\right),\end{equation}
and the vorticity, $\omega$, is 
\begin{equation}\omega=-\left(\frac{\partial^2}{\partial
x^2}+\frac{\partial^2}{\partial z^2}\right)\psi.\label{PsiEqn}\end{equation}
The above sign conventions (required by a referee of KD) result in $\omega$
being the component of vorticity in the {\em negative} $y$-direction.

The length scale used for the nondimensionalisation is $k^{-1}$, where
the assumed periodicity of the vortices in the $x$-direction is
$2\upi/k$. The stream function is scaled with the kinematic
viscosity, $\nu$.  The nondimensional parameter 
\begin{equation}\lambda=\frac{A'}{\nu k^2}\end{equation} 
is a measure of the strength of the converging flow compared to the rate of viscous
dissipation.

We can expand the periodic solutions in terms of Fourier series in the $x$-component:
\begin{subequations}
\begin{equation}\omega(x,z)=\sum^\infty_{n=0}a_n(z)\cos{nx}+b_n(z)\sin{nx},\label{FourierA}\end{equation}
\begin{equation}\psi(x,z)=\sum^\infty_{n=0}c_n(z)\cos{nx}+d_n(z)\sin{nx}.\label{FourierB}\end{equation}
\end{subequations}
Unlike KD, we allow the possibility for all the functions $a_n(z)$, $b_n(z)$, $c_n(z)$ and $d_n(z)$ to be non-trivial. These Fourier series are substituted into (\ref{OmEqn}) and (\ref{PsiEqn}), giving rise to an infinite set of ordinary differential equations
\begin{subequations}
\begin{equation}a_n''+\lambda z a_n' +(\lambda-n^2)a_n=J_{n}(\psi,\omega),\label{SystemEqnStart}\end{equation}
\begin{equation}b_n''+\lambda z b_n' +(\lambda-n^2)b_n=K_{n}(\psi,\omega),\end{equation}
\begin{equation}c_n''-n^2c_n=-a_n,\end{equation}
\begin{equation}d_n''-n^2d_n=-b_n,\label{SystemEqnEnd}\end{equation}
\end{subequations}
where $J_{n}(\psi,\omega)$ and $K_{n}(\psi,\omega)$ are the appropriate nonlinear terms. This system of equations was truncated as before, and 
solved using a fourth order Runge-Kutta scheme. As in KD we will require 
all perturbation velocities to decay to 0 as $z\rightarrow\pm\infty$. 
For large $|z|$ a shooting scheme was used to ensure that the 
functions decayed with the 
correct asymptotic behaviour (see the next section).

In KD it was found that satisfactory convergence could be obtained if
the system of equations was truncated after $n=12$. For some of the
examples in this current work it was necessary to continue up to
$n=48$.  One test of convergence was to examine the amplitudes of the vorticity at the cores and in the far field. The ratio of these would only converge once the accuracy was sufficient to resolve some of the
finer scale structures in the vorticity seen in some of the large
amplitude solutions discussed in section 5.
This linkage between the amplitude of the vortices and the far field was not investigated in KD, hence a lower resolution was adequate there.

\section{Large $|z|$ behaviour}

The behaviour of the decaying vorticity as $z\rightarrow\pm\infty$
plays a crucial r\^ole in the form of the periodic vortices that was
not appreciated in KD. The large $|z|$ behaviour of
the vorticity tail for the Fourier mode proportional to $\cos nx $ (or
$\sin nx$), where the nonlinear terms can be neglected, satisfies the
linearised equation
\begin{equation}
a_n''+\lambda za_n'+(\lambda-n^2)a_n=0.\label{LinMode}
\end{equation}
This equation has two possible sorts of leading order behaviour in
the large $|z|$ limit. As $z\rightarrow\infty$ either
\begin{equation}
a_n\sim Bz^{-(\lambda-n^2)/\lambda}
\label{SlowDecay}\end{equation}
or
\begin{equation}
a_n\sim C\int_z^\infty {\rm e}^{-\lambda z'^2/2}\,{\rm d}z'.
\end{equation}
The second of these is the behaviour found in the similarity solutions
of Lin \& Corcos (1984). The behaviour in (\ref{SlowDecay}) may at
first seem unexpected. However, if one considers the behaviour of
vorticity with this assumed periodicity in the $x$-direction, and no
variation in the $z$ direction then it will be found to grow
exponentially in time due to vortex stretching as
\begin{equation}
\omega\propto\exp(\lambda-n^2)t.
\end{equation}
Far from $z=0$ the $z$ variation of the vorticity is weak, and so a
parcel of vorticity will indeed grow with approximately this growth
rate.  However, the parcel is also being advected towards the plane
$z=0$ with velocity $-\lambda z$, and so its distance from the plane
$z=0$ will vary with time as $\exp(-\lambda t)$. Thus the vorticity as
a function of $z$ will behave as
\begin{equation}
\omega\propto z^{-(\lambda-n^2)/\lambda}.
\end{equation}
This is the behaviour found in (\ref{SlowDecay}). Thus the algebraic 
tail to the vorticity is a
manifestation of the vorticity growing exponentially as it approaches
$z=0$.  As the vorticity approaches the plane of symmetry the length
scale in the $z$-direction decreases thus increasing the rate of viscous
dissipation, and limiting further growth.

If we look at the equations for linear perturbations we find that
non-trivial solutions can have the super-exponential decay for either
positive or negative $z$ but not both. This can be seen by integrating
(\ref{LinMode}) between two limits, say $z=\alpha$ and $z=\beta$, to
give
\begin{equation}
a_n'(\beta)-a_n'(\alpha)+\lambda\beta a_n(\beta)-\lambda \alpha a_n(\alpha)-n^2\int^\beta_\alpha a_n(z)\,{\rm d}z=0.\label{Integral}
\end{equation}
From this equation we can deduce a series of results.

The first result is that no solutions exist with more than one zero.
For example, if we assume that we have a solution which is positive
between two zeros at $\alpha$ and $\beta$, with $\alpha<\beta$, then
the first two terms cannot be positive, and the next two terms are zero.
There are no positive contributions to balance the final negative
integral term. Hence we conclude it is not possible for there
to be a function $a_n(z)$ which satisfies (\ref{LinMode}) and has two
zeros.

The second result is that we cannot have a zero and superexponential
decay in either direction. For example if we have a zero at $z=\alpha$
with $a_n(z)>0$ for $z>\alpha$ and assume fast decay for large
positive $z$, then looking at the limit as $\beta\rightarrow\infty$ in
(\ref{Integral}) would give the first and third terms both tending to
0 rapidly, the second term would be non-positive, and the fourth term zero.
Again there is nothing to balance the negative contribution from the
integral. This result can be extended to both directions, and so we
conclude that if $a_n(z)$ has a zero then it must decay algebraically
in both directions.

The last result is that we cannot have a one-signed function with two
rapidly decaying tails. If we had a positive function which behaved 
in this way, and we looked at the limits as
$\alpha\rightarrow-\infty$ and $\beta\rightarrow\infty$, then the first
four terms of (\ref{Integral}) must all tend to 0, and so the integral
of $a_n(z)$ must also be zero. This is not possible for a positive
function, so we conclude that we must have at least one algebraically
decaying tail.

The behaviour described above is useful in understanding the possible
solutions to the governing equations. We see that without a supply of
vorticity being swept towards the plane $z=0$ there would be no
steady-state vortices. A similarity solution for the case with rapidly
decaying vorticity in both directions was found by Lin \& Corcos
(1984). Their solutions were all exponentially decaying with time. We
can regard the vortices observed near the plane $z=0$ as being the
result of the vorticity that is supplied at some point a long way from
this plane.  In KD it was assumed that the fundamental mode of the
vorticity was an even function in $z$ and the numerical solutions were
found by imposing the maximum value of the stream function at the
origin and allowing vorticity in the algebraic tail of the fundamental
mode to decay as appropriate.  This choice was made based on the
original expected form of the vortices.  This assumption gave rise to
solutions where the vorticity tail from negative $z$ and that from
positive $z$ were of the same magnitude and with incoming positive
vorticity  aligned. It would be equally possible to have the vorticity
tails of opposite signs, in which case the fundamental mode would still
consist of the same Fourier modes, but the fundamental modes of the
vorticity and stream function would be odd functions in $z$.
Alternatively there could be considered to be an arbitrary shift in the
alignment between the incoming vorticity.  These are some of the
additional possibilities which will be examined in the next section.

\section{Further periodic vortices}

In general it is possible to specify the size of the algebraically
decaying tails of each mode of vorticity that satisfies the condition
$n^2<\lambda$ arbitrarily as $z\rightarrow\pm\infty$.  There is a
vast potential range of such solutions. However, we will restrict
ourselves to situations where only the fundamental modes $\cos x$ and
$\sin x$ have non-zero tails coming in from infinity. It is these modes
which have the greatest amplification of the vorticity as it is swept 
towards $z=0$ and so would be expected to dominate.
We can write the leading order behaviours of the two tails as
\begin{subequations}
\begin{equation}
\omega\sim-A_+z^{-(\lambda-1)/\lambda}\cos(x-\phi_+)\qquad\hbox{as}\qquad z\rightarrow+\infty,
\end{equation}
\begin{equation}
\omega\sim-A_-|z|^{-(\lambda-1)/\lambda}\cos(x-\phi_-)\qquad\hbox{as}\qquad z\rightarrow-\infty.
\end{equation}
\end{subequations}
We will refer to the coefficients $A_+$ and $A_-$ as the amplitudes of
the respective algebraic tails, and $\Phi=(\phi_+-\phi_-)$ as the phase
difference between the tails. There is no requirement here that the
amplitudes are both positive. This means that these quantities are not
uniquely defined, but the convenience of this extra degree of freedom
will be apparent in the subsequent subsections.

In the following subsections we will look at three of these families of
solutions. In each case we will restrict ourselves to the case of
$\lambda=6$. This is the same choice used in KD to look at
the variation of amplitudes of the vortices presented there.

In the first section we will look at solutions where the incoming
vorticity is aligned and of equal magnitude, but of opposite sign. In the
next we will still keep the incoming vorticity aligned but vary the
relative strengths of the incoming vorticity. Lastly we will keep the
amplitudes of the incoming vorticity the same from both directions, but
allow an arbitrary phase shift between the incoming streams.

\subsection{Vorticity tails aligned with opposite magnitudes}

The flows found here have perturbation stream functions and vorticity
that can be described in terms of the same Fourier modes as in KD:
\begin{subequations}
\begin{equation}\omega(x,z)=\sum^\infty_{n=1}a_n(z)\cos{(2n-1)x}+b_n(z)\sin{2nx},\end{equation}
\begin{equation}\psi(x,z)=\sum^\infty_{n=1}c_n(z)\cos{(2n-1)x}+d_n(z)\sin{2nx}.\end{equation}
\end{subequations}
With only the $\cos x$ mode decaying algebraically as $z\rightarrow\pm\infty$ all the Fourier modes omitted from the general expansion (\ref{FourierA}--\ref{FourierB}) are identically zero (see KD).
This time the functions $a_n(z)$ and $c_n(z)$ are odd while the
functions $b_n(z)$ and $d_n(z)$ are even. We will have equal tail
amplitudes ($A_+=A_-$), and a phase difference $\Phi=\upi$ (or
alternatively $A_+=-A_-$ and $\Phi=0$). As in KD we will define the
amplitude of the solutions as the maximum value of the stream function
$\psi(x,z)$. This maximum may no longer be situated at the origin. A
typical sequence of flows is shown in figure 1. These are calculated for
$\lambda =6$. Initially the perturbation vorticity is not sufficiently
strong to induce a swirling motion in the total flow. As the amplitude
increases, pairs of vortices appear. These move further away from the
plane $z=0$ as the swirling motion becomes stronger.

As the amplitude of the odd solutions increases the vortices appear to
pair up and then move away from the plane of symmetry. The mechanism
behind this behaviour is outlined in figure 2. The influence of each
vortex's mirror image in the plane $z=0$ is to induce a horizontal
displacement in the vortices which alternates in direction between adjacent
vertical pairs (figure 2(a)). The vortices in line above the $z$-axis
are now no longer evenly spaced, and each vortex is more strongly
influenced by its nearest lateral neighbour, and experiences a net
tendency to move into the incoming flow (figure 2(b)).  This tendency
to move away from the plane of symmetry is limited by the stronger
inflow as $|z|$ increases (figure 2(c)).

The linear analysis for the structure of the vortices in KD in the
large $\lambda$ limit is readily adapted to these solutions, and also
to those of the next two sections. The analysis is not repeated here.

\subsection{Vorticity tails aligned with arbitrary magnitudes}

If we fix the phase difference between the tails to be zero, i.e.
$\Phi=0$, but allow  the amplitude of the tails to take arbitrary
values we are able to incorporate both the solutions of KD and those of
the previous section.  The same Fourier modes as before will be
non-zero, but now there is no restriction of the symmetry of the
functions $a_n(z)$, $b_n(z)$, $c_n(z)$ and $d_n(z)$.  A sequence of
such flows is shown in figure 3. These flows all have the amplitude of
the tails from large positive $z$ fixed at $A_+=10$, but with the
opposing tails having amplitudes ranging from $A_-=-10$ to $A_-=10$.
Initially two pairs of vortices are visible such as those seen in the
previous section. As the magnitude of the lower vorticity tail
decreases, the lower pair rapidly cease to exist, although the
associated local maxima and minima of vorticity continue to exist
for some time. As $A_-$ increases the remaining pair of vortices grow
until they eventually recover the form of the vortices investigated in
KD.

A feature that is clear in the sequence in figure 3 is the great
disparity between the strength of the vortices when the tails have the
same signs and when they have opposite signs. When the tails have
opposite signs there is a region of enhanced viscous dissipation near
$z=0$ where the regions of opposite vorticity are being swept together.
This region is not present when the tails are of the same sign.

The double vortices of the previous subsection are quickly swamped as
the amplitude of the lower tail increases, although the  vorticity only
becomes single-signed in each half-period when $A_-=0$. For the
majority of possible values of the lower tail the vortices closely
resemble those associated with tails of the same magnitude and sign.
The most noticeable difference is a slight shifting of the vortex cores
above the plane $z=0$. This displacement diminishes as $A_-\rightarrow
A_+$. If we plot the maximum value of $\psi$ as a function of $A_-$ 
(figure 4) we see that the curve is roughly linear for much of the
interval. There is an almost linear relationship between the amplitude 
and the sum of the tails for positive $A_-$. The relationship between 
the vortex amplitudes and the tails will be examined in more detail 
in section 5.

\subsection{Vorticity tails with equal magnitudes and arbitrary phase}

The last class of solutions that we will examine here are those with
tails of the same magnitude but with no restriction on their
alignment.  Thus the possible extremes are again when the tails are
alinged and of opposite sign, and when they are aligned and of the same
sign. A sequence showing the transition between these extremes is shown
in figure 5 for $\lambda=6$ and a tail strength of $A_+=A_-=10$. This
sequence shows a monotonically increasing amplitude for the vortices,
but the phase difference is not monotonic.  It decreases from the
initial value $\Phi=\upi$, through 0 to a small negative number,
$\Phi=-0.081\upi$.  The phase difference then increases again, passing
through 0 to $\Phi=0.595\upi$ before decreasing again to 0 where the
sequence ends with the same solution as found before.  The curves of
the maximum and minimum values of $\psi$ are shown in figure 6 along
with the corresponding curves for $\Phi$ ranging from $-\upi$ to 0.
These curves consist of three sets of branches of solutions: the lower
branches  connected to the solutions which exist when $\Phi=\pm\upi$, 
the upper branch
connected to the KD solution for $\Phi=0$ and the middle branches that
connect the ends of these two branches.  We see that for some values of
$\Phi$ multiple solutions are possible. For example,  for $\Phi=0$
there are 5 possible flows: the symmetric KD solution found previously
on the upper branch, and a pair of solutions on each of the two
other sets of branches of solutions.  In each of these latter pairs the
solutions are mirror images of each other. The width of the upper
branch increases with increasing $\lambda$. The upper branch for
$\lambda=8$ and tail magnitude of 10 is also included in figure 6. This
branch of solutions exists for almost the entire range of values of
$\Phi$.

The large amplitude of the upper branch is not strongly affected by the
phase difference between the tails. The incoming vorticity is still
swept into the vortices, but will take a longer or shorter route
depending on the sense of rotation of the vortices and the phase
difference. The vortices where this route is shorter tend to be
relatively strong, while the counter rotating vortices with the longer
path are relatively weak.

It may be anticipated that the upper and lower branches of solutions
are stable and the middle branch is unstable. However, no stability
analysis of these solutions has yet been performed.

\section{Link between tail and vortex amplitudes}

One of the features that we can investigate is the maximum intensity of
the vortices compared to the magnitude of the algebraic vorticity
tails. This is illustrated in figure 7 for the solutions with both odd
and even fundamental modes with $\lambda=6$. There is a small degree of
hysteresis in the odd case, but not in the even one. However there is a
significant jump in the rate of increase in amplitude with tail
strength when $A_+$ is around 2. Also shown in this figure is the
corresponding curve for the even mode with $\lambda=8$. Here there is
clear hysteresis for the even mode. The details of the causes of this
jump differ between the odd and even cases.

In the case of the solutions with the even fundamental mode (the KD
solutions) the cause of the jump is that as the amplitude of the
instabilities increases then the distribution of the vorticity
changes.  Instead of being confined to a region of thickness
$\lambda^{-1/2}$ as shown by the linear large-$\lambda$ asymptotics of
KD, a region more extensive in the $x$-direction than the
$z$-direction, the vorticity takes a more elliptical form with its
major axis parallel to the $z$-axis. This results in the typical 
length scale for variations in the vorticity in the $z$-direction 
increasing, reducing the dissipation.  This transition is shown by 
the vorticity distributions displayed in figure 8.  These correspond to 
the flows shown in figures 5 and 6 of KD.

When the tails of vorticity are of the same magnitude but of opposite
signs, as in section 4.2, the argument is slightly different. Again the
initial configuration of vorticity conforms to that of the large
$\lambda$ asymptotics, but the rearrangement of vorticity has much to
do with the vortex interactions discussed in section 4.2. These
interactions move the vortices away from each other and thus
significantly reduce the dissipation. In the KD case the vortices are
unable to move away from each other. This extra mechanism allows for a
more effective reduction in dissipation, thus the  hysteresis is
observed for lower values of $\lambda$.

The  lower end of all the curves in figure 7 have slope 1. In this
region the solutions are essentially linear and so this is the
behaviour one would expect. However, the slope on the upper part of the
curve for the solutions with the even fundamental mode (the solutions
of KD) is less than 1, thus a doubling of the amplitude of the tail
produces a smaller increase in the maximum of the stream function.
Along this branch of solutions the distribution of the vorticity near
the core of each vortex is well described by the large amplitude
asymptotics of KD. However the asymptotic analysis presented there did
not take into account any linkage between the cores of the vortices and
the algebraically decaying vorticity tails. This linkage is a complex
one.  In figure 9 is shown the vorticity distribution for a KD vortex
with $\lambda=6$ and amplitude 5000. It shows how the tail of vorticity
behaves as it approaches the large vortex core. There are three
important regions. In addition to the vortex core, there is the region
of algebraic decay for larger vales of $z$. This extends almost all the
way down to the the stagnation point in the flow above the region
between the left pair of vortices.  The vorticity then enters into a
transition region where it is swept towards the edge of the vortex
core. In the process it is squeezed into a narrow region which leads to
enhanced lateral dissipation, leading to a reduction in the magnitude
of the vorticity as it is swept towards the vortex. This vorticity then
merges with the vortex core.

As the amplitude of the vortices grows, the position of
the stagnation point moves further away. Thus for a given amplitude of
tail the magnitude of the vorticity as it enters the transition region 
will be lower. In addition the distance from this stagnation point to
the vortex will be higher, giving more opportunity for viscous
dissipation. These effects will lead to the ratio of the vorticity
arriving in the core to the amplitude of the tail decreasing as the
amplitude of the vortices increases.

One further feature of the distribution of the vorticity in figure 9
should be emphasised, and that is the large difference between the
amplitude of the vorticity in the vortex cores and the vorticity being
swept in from the tails. The magnitude of the vorticity in the core of
this example
exceeds 12\,000, while the local maximum in the incoming vorticity is just
over 30. This weak feed of vorticity is required to maintain the steady
vortices.  If this supply was cut off the effect would be that the
magnitude of the vortex would decay, but it may be anticipated that the
rate of decay would be quite small.

The details of how all these regions link up is not clear. For larger
values of $\lambda$ the structure of the vortex cores is similar to
those found by Moffatt, Kida \& Ohkitani (1993) in the investigation of
single vortices in a more general range of stagnation point flows. This
similarity is especially strong for larger values of $\lambda$. One
example of this similarity is shown in figure 10. The dissipation rate
in the vortex cores displays the same structures that they found with
their asymptotic analysis, and which have been observed elsewhere (for example, Kida \& Ohkitani, 1992). In their analysis they are unable as yet to determine the
structure of the decaying vorticity far from the vortex core. Their
problem does not have the added complication of incoming vorticity that
is present here.

\section{Discussions}

The understanding of how the steady-state periodic vortices in a
stagnation point flow of Kerr \& Dold (1994) are essentially driven by
the inflow of vorticity from the far field has led to the discovery of
further steady-state periodic vortex solutions. These do not have the
original assumed symmetry of KD.  We have only presented three new
families of solutions here, and concentrated on one choice of
$\lambda$. It is clear that there is a vast range of further solutions,
many of which have not been investigated. However, the ones presented
display many of the important features that would be observed in these
other solutions.

In all these examples the vorticity is not confined to a finite region
around the core of the vortices as in the case of the Burgers vortex
(Burgers, 1948). For steady periodic vortices to exist there is a
requirement for a supply of upstream vorticity.  The need for this
upstream supply of vorticity for steady vortices is not necessarily an
onerous condition. The ratio of the magnitude of the vorticity at the
vortex cores to that at even a moderate distance away from the plane $z=0$ can
be very large. Thus a seemingly weak level of disturbance in the
upstream conditions in a typical experiment may result in the
generation of large amplitude vortices.

One major area that has not yet been addressed is how these vortices
would come into being. The temporal evolution of fully nonlinear vortices is
beyond the scope of this paper. However the growth of linear
instabilities is a simple extension of the current work. If one
allowed for exponentially growing solutions with a nondimensional
growth rate $\sigma$, then the equations for each mode become
\begin{equation}
a_n''+\lambda za_n'+(\lambda-\sigma-n^2)a_n=0.\label{GrowMode}
\end{equation}
The analysis of this equation follows that of section 2, and all
conclusions reached there will carry through for solutions with
$\sigma$ positive with minor changes. Thus exponentially growing
solutions to the linear equations can only exist if there is an
appropriate supply of vorticity from afar. From this we can see that
growing solutions will be a response to an increasing level of
vorticity being supplied from some upstream source. If the far-field
evolved in a  way that resulted in the supply of vorticity increasing
in some other way then the variables will not be separable and the 
appropriate partial differential equations would have to be solved.

The need for algebraic tails to the vorticity raises potential problems
in situations where vortices are observed to grow in what are
essentially stagnation point flows. One such case is the growth of
secondary instabilities to Kelvin-Helmholtz rolls in a shear layer as
discussed in KD. These secondary instabilities bear a close resemblance
to the vortices discussed here and in KD. This leads one to expect that
these instabilities require an increasing supply of vorticity from
upstream.  An investigation of this region for exponentially growing
instabilities that  has strictly localised vorticity is doomed to
failure.  This need for an upstream supply of  vorticity may be
satisfied by vorticity being supplied by the vortices in adjacent
regions of stagnation point flow in the series  of Kelvin-Helmholtz
rolls. This possibility is investigated in Kerr (1997).

If one considers the kinetic energy of these disturbances a distinct
difference is seen between these perturbations and those of Craik \&
Criminale (1986) or the fundamental modes of Lagnado, Phan-Thien \&
Leal (1984). If an energy per period per unit length in the
$y$-direction of the perturbations is defined
\begin{equation} E=\frac{1}{2}\int_{-\infty}^\infty\int_{-\upi}^\upi|{\bf
u}|^2\,{\rm d}x\,{\rm d}z, \end{equation}
then, with appropriate rescalings, the
energy of the disturbances in the other works is infinite. In the
examples here and in KD the velocities decay as $|z|^{-1+1/\lambda}$.
Thus the integral of the energy will converge provided $\lambda>2$.
This is only slightly more restrictive than the condition for decaying
solutions to exist, $\lambda>1$. When $\lambda>2$ this clearly
indicates that the disturbances are extracting energy from the
stagnation point flow, and are not just redistributing an infinite
supply of energy.

The understanding of the periodic vortices in a stagnation point flow 
presented here and in Kerr \& Dold (1994) has benefited from the 
appreciation of the requirement for algebraic tails to the vorticity 
in the upstream direction. This may seem to be a restriction 
in their applicability to real flows, however it could be considered 
to be a guide to where one should look for instability mechanisms. 
If these vortices are considered to be the signature of an upstream 
supply of vorticity then that is where one should look.

\typeout{********************************}
\typeout{* Acknowledgement to Bill Dold *}
\typeout{********************************}

\newpage
\noindent{\Large \bf Figure Captions}

\bigskip
\begin{list}{}{\setlength{\rightmargin}{0pt}
               \setlength{\leftmargin}{0pt} 
               \setlength{\itemindent}{0pt} 
               \setlength{\parsep}{0pt} 
               \setlength{\itemsep}{8pt} }
\item{\sc Figure 1.} Sequence of flows with $A_+=-A_-$ and $\Phi=0$. The
top row shows stream lines in the $x$--$z$ plane for the full flow, the
second row $\omega$ and the last $\psi$. In the last two rows the
contour spacing is a sixth on the maximum modulus of $\psi$ and
$\omega$ respectively. The values of $A_+$ and the corresponding
maxima of $\psi$ and $\omega$ are
(a) 2, 1.17, 3.11,  
(b) 20, 13.04, 37.30,
(c) 30, 27.39, 73.93, 
(d) 35.6, 62.60, 161.3,
(e) 35.6, 94.07, 238.4, 
(f) 34.6, 164.4, 408.9,
(g) 40,  568.7, 1389.

\item{\sc Figure 2.} Schematic diagram showing the effect of vortex
interactions. (a) Initially vertical pairs of counter-rotating vortices
interact to move sideways. (b) Horizontal rows no longer evenly spaced
with the influence of nearest neighbours tending to make vortices move
away from $z=0$. (c) Pairs of vortices come to rest in region where the
inward flow is sufficiently strong.

\item{\sc Figure 3.} Sequence of flows with $A_+=10$ and $A_-$ ranging
from $-10$ to $+10$, with $\Phi=0$. The top row shows stream lines in
the $x$--$z$ plane for the full flow, the second row $\omega$ and the
last $\psi$. In the last two rows the contour spacing is a sixth on the
maximum modulus of $\psi$ and $\omega$ respectively. The values of
$A_-$ and the corresponding maxima of $\psi$ and $\omega$ are
(a) -10, 5.90, 16.49, 
(b) -9,  7.28, 19.53,
(c) -7, 12.14, 30.22,   
(d) -5, 36.33, 89.08,
(e) 0, 244.3, 594.3,
(f) 5, 411.9, 1001,
(g) 10, 562.6, 1368.

\item{\sc Figure 4.} Graph of the maximum value of $\psi$ of the
vortices as a function of $A_-$, with $A_+=10$ and $\lambda=6$.

\item{\sc Figure 5.} Sequence of flows with $A_+=A_-$ for a range of
$\Phi$. The top row shows stream lines in the $x$--$z$ plane for the
full flow, the second row $\omega$ and the last $\psi$. In the last two
rows the contour spacing is a sixth on the maximum modulus of $\psi$
and $\omega$ respectively. The values of $\Phi$ and the corresponding
maxima are
(a) $\upi$ -10, 5.90, 16.49,
(b) $0.9\upi$, 6.16, 17.21,
(c) $0.6\upi$, 21.30, 48.69,
(d) $-0.075\upi$, 66.84, 152.0,
(e) $0.25\upi$, 113.6, 251.9,
(f) $0.595\upi$, 175.8, 403.5,
(g) 0, 562.6, 1368.

\item{\sc Figure 6.} Graph of the sizes of the maxima (solid line)
and minima (dashed line) of $\psi$  for $\lambda=6$, $A_+=A_-=10$ and
$\Phi$ ranging between $-\upi$ and $\upi$. Superimposed on each is the
upper part of the corresponding curves for $\lambda=8$ (dotted line).

\item{\sc Figure 7.} Graphs of the maximum values of $\psi$ for
vortices  as a function of the tail strength $A_+$. The solid and
dotted curves are for KD vortices ($A_+=A_-$, $\Phi=0$) with
$\lambda=6$ and $\lambda=8$ respectively, while the dashed curve is for
the vortices described in section 4.1 ($A_+=-A_-$, $\Phi=0$) with
$\lambda=6$.

\item{\sc Figure 8.} Vorticity plots for the vortices corresponding to
figures 5 and 6 of KD. These vortices have amplitudes (a) 2.5, (b) 10,
(c) 40 and (d) 160. The contour spacing is a tenth of the maximum value
of $\omega$. In each case this maximum is (a) 5.696, (b) 23.16, (c)
75.13, and  (d) 388.8.

\item{\sc Figure 9.} Plots of the (a) streamlines and (b) vorticity of
the upper half of vortices with $\lambda=6$ and amplitude 5000. The
contour intervals in (b) are 1216 near the vortex cores, and 10
elsewhere. 

\item{\sc Figure 10.} Dissipation for KD vortex with $\lambda=50$ and
amplitude 80. 

\end{list}  

\newpage

\begin{figure}
\centerline{\includegraphics[trim=5mm 10mm 25mm 5mm,width=5.3in]{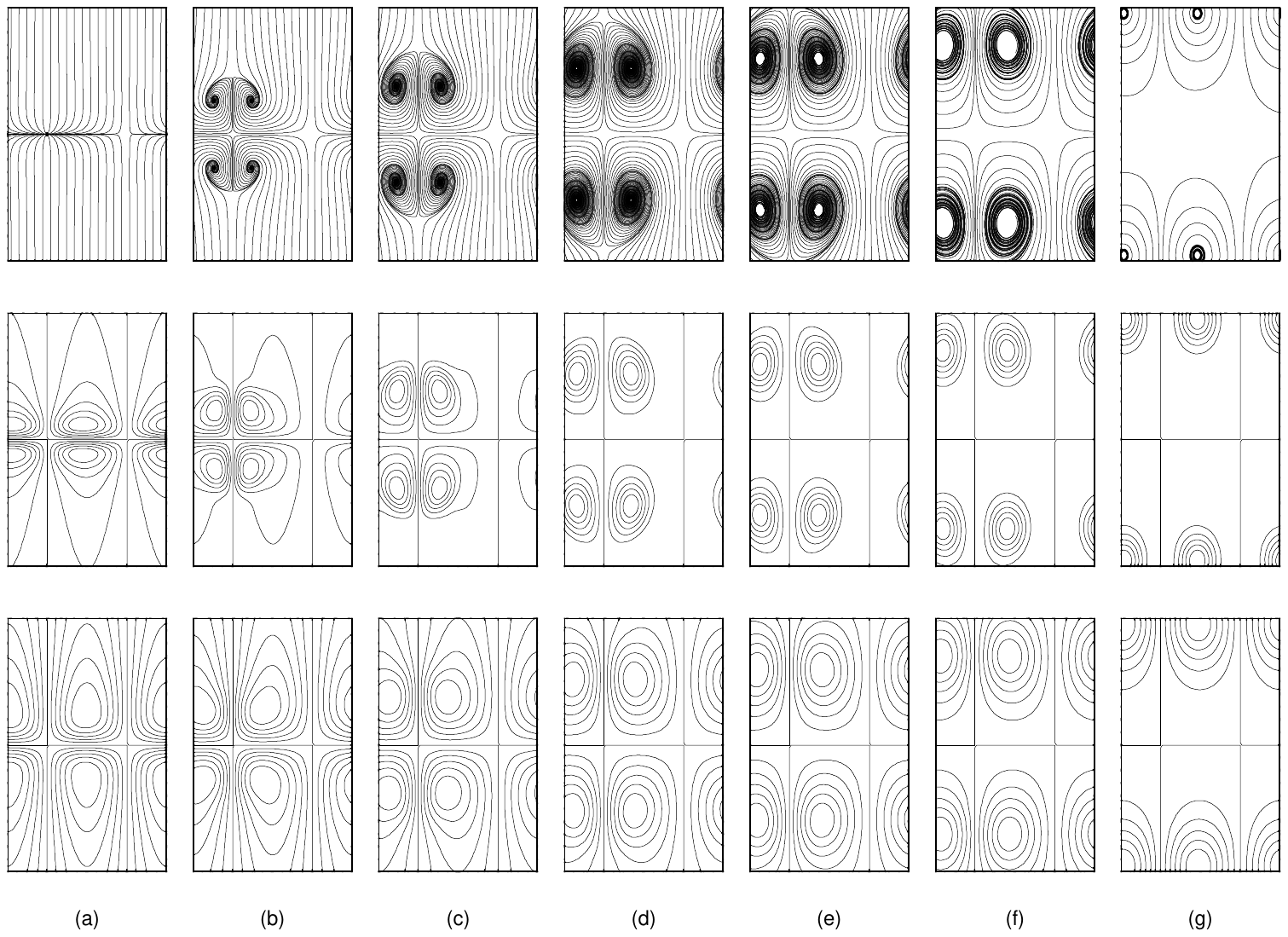} }
\caption{Sequence of flows with $A_+=-A_-$ and $\Phi=0$. The
top row shows streamlines in the $x$--$z$ plane for the full flow, the
second row $\omega$ and the last $\psi$. In the last two rows the
contour spacing is a sixth of the maximum modulus of $\psi$ and
$\omega$ respectively. The values of $A_+$ and the corresponding
maxima of $\psi$ and $\omega$ are
({\it a}) 2, 1.17, 3.11,  
({\it b}) 20, 13.04, 37.30,
({\it c}) 30, 27.39, 73.93, 
({\it d}) 35.6, 62.60, 161.3,
({\it e}) 35.6, 94.07, 238.4, 
({\it f}) 34.6, 164.4, 408.9,
({\it g}) 40,  568.7, 1389.}
\label{Fig1}
\end{figure}

\begin{figure}
\centerline{\includegraphics[width=2.4in]{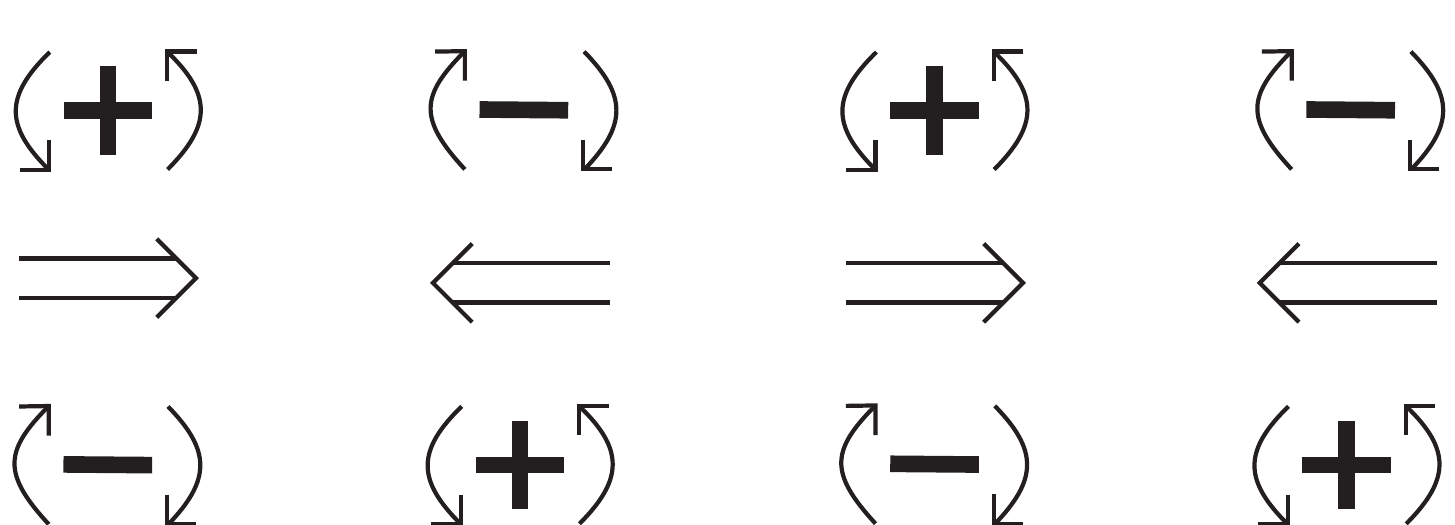}}
\centerline{({\it a})}
\centerline{\includegraphics[width=2.4in]{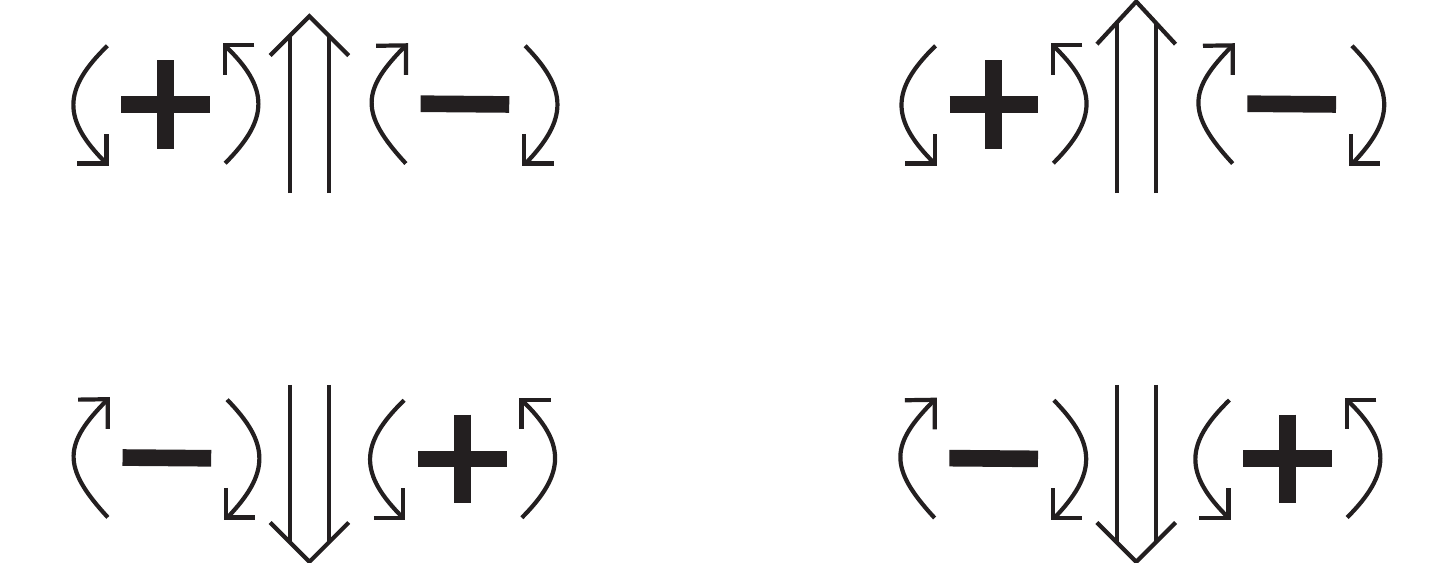}}
\centerline{({\it b})}
\centerline{\includegraphics[width=2.4in]{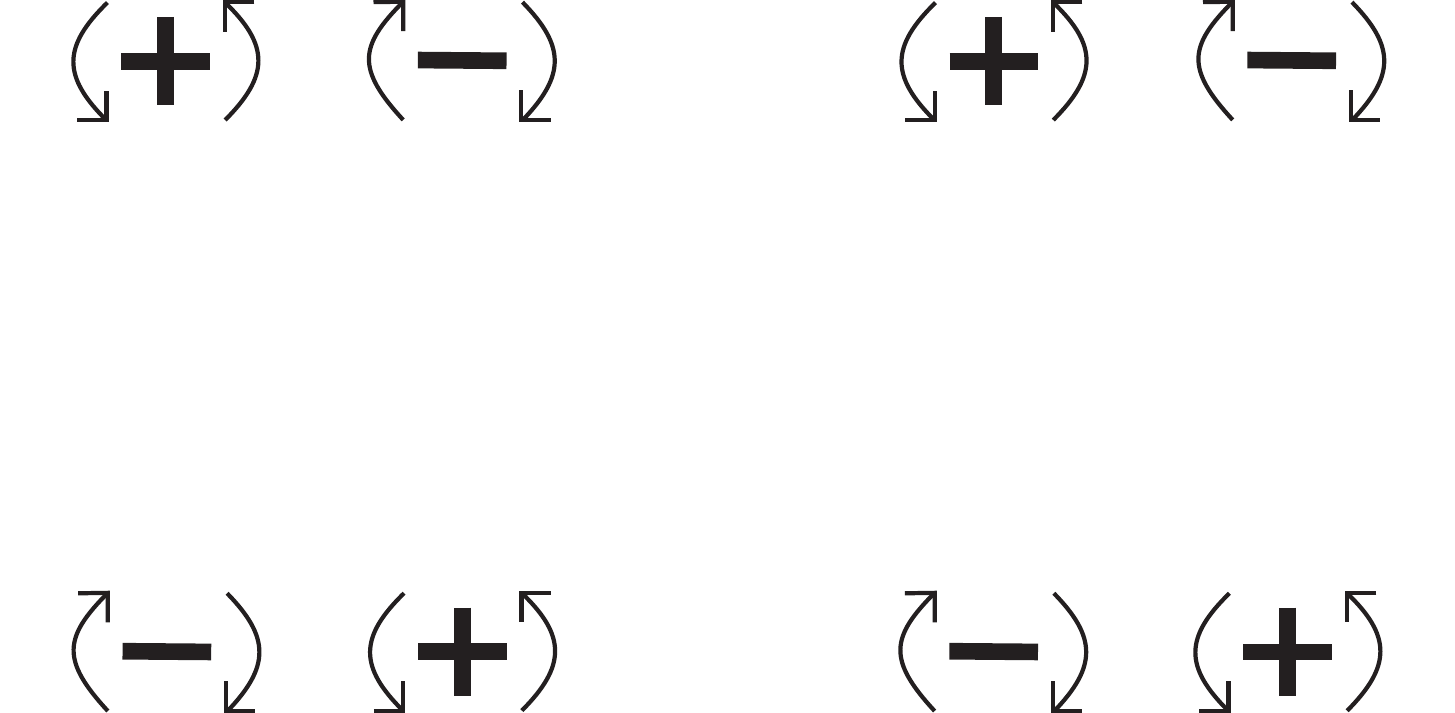}}
\centerline{({\it c})}
\caption{Schematic diagram showing the effect of vortex
interactions. ({\it a}) Initially vertical pairs of counter-rotating vortices
interact to move sideways. ({\it b}) Horizontal rows no longer evenly spaced
with the influence of nearest neighbours tending to make vortices move
away from $z=0$. ({\it c}) Pairs of vortices come to rest in region where the
inward flow is sufficiently strong.}
\label{Fig2}
\end{figure}

\begin{figure}
\centerline{\includegraphics[trim=5mm 10mm 25mm 5mm,width=5.3in]{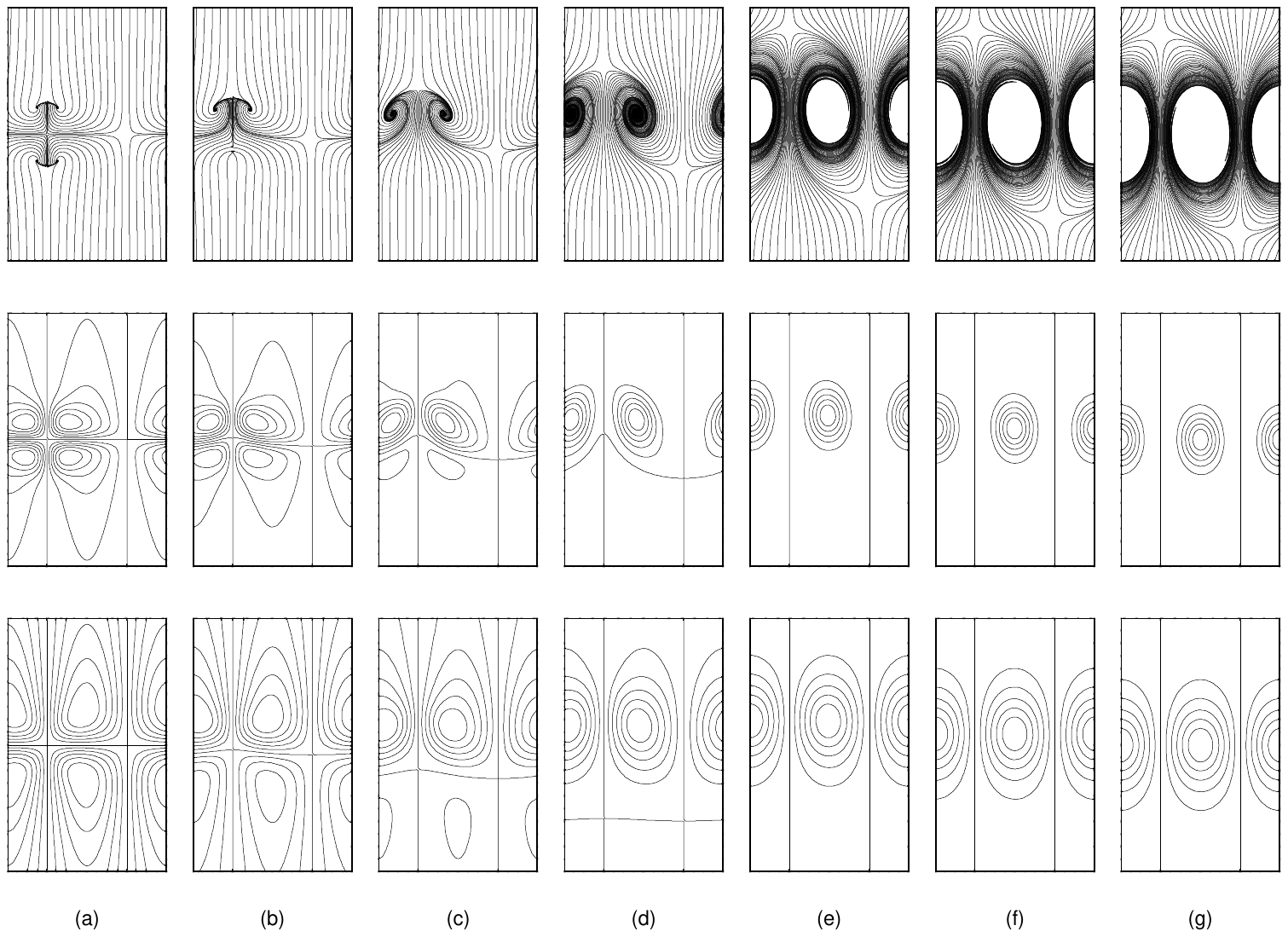} }
\caption{Sequence of flows with $A_+=10$ and $A_-$ between
 $-10$ and $+10$, with $\Phi=0$. The top row shows streamlines in
the $x$--$z$ plane for the full flow, the second row $\omega$ and the
last $\psi$. In the last two rows the contour spacing is a sixth of the
maximum modulus of $\psi$ and $\omega$ respectively. The values of
$A_-$ and the corresponding maxima of $\psi$ and $\omega$ are
({\it a}) -10, 5.90, 16.49, 
({\it b}) -9,  7.28, 19.53,
({\it c}) -7, 12.14, 30.22,   
({\it d}) -5, 36.33, 89.08,
({\it e}) 0, 244.3, 594.3,
({\it f}) 5, 411.9, 1001,
({\it g}) 10, 562.6, 1368.}
\label{Fig3}
\end{figure}

\begin{figure}
\centerline{\includegraphics[trim=10mm 10mm 10mm 130mm,width=3.0in]{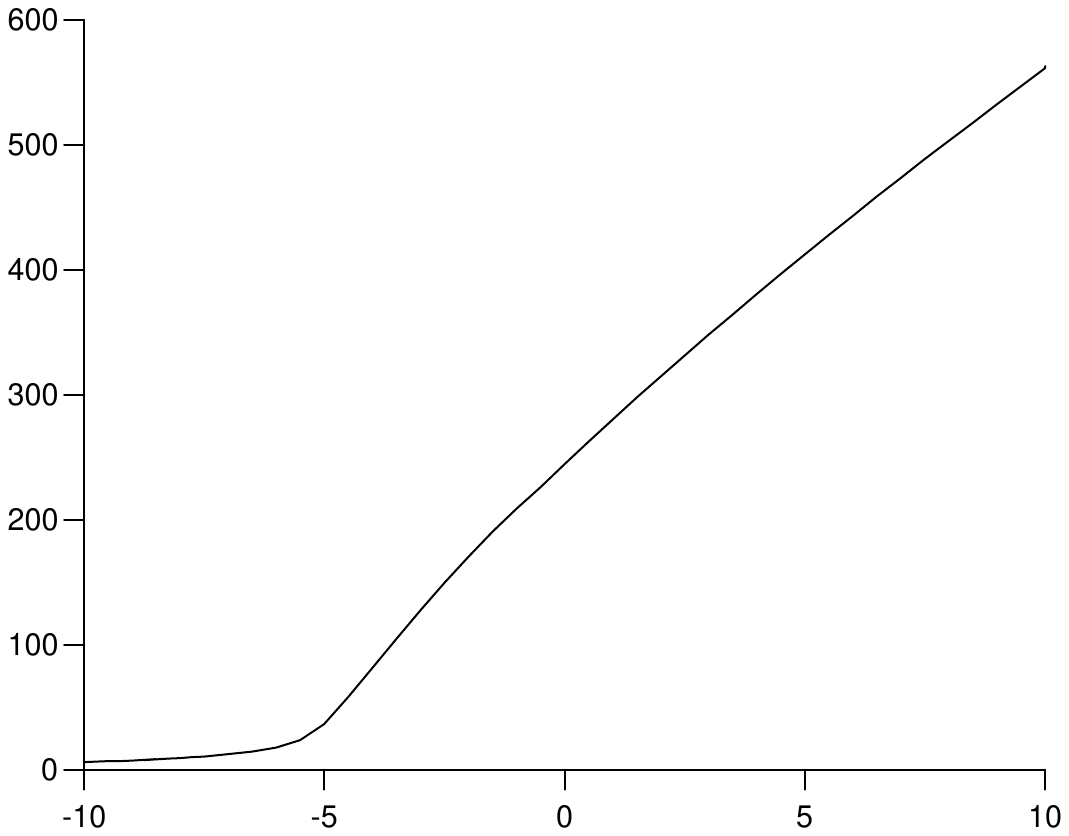} }
\caption{Graph of the maximum value of $\psi$ of the
vortices as a function of $A_-$, with $A_+=10$ and $\lambda=6$.}
\label{Fig4}
\end{figure}

\begin{figure}
\centerline{\includegraphics[trim=5mm 10mm 25mm 5mm,width=5.3in]{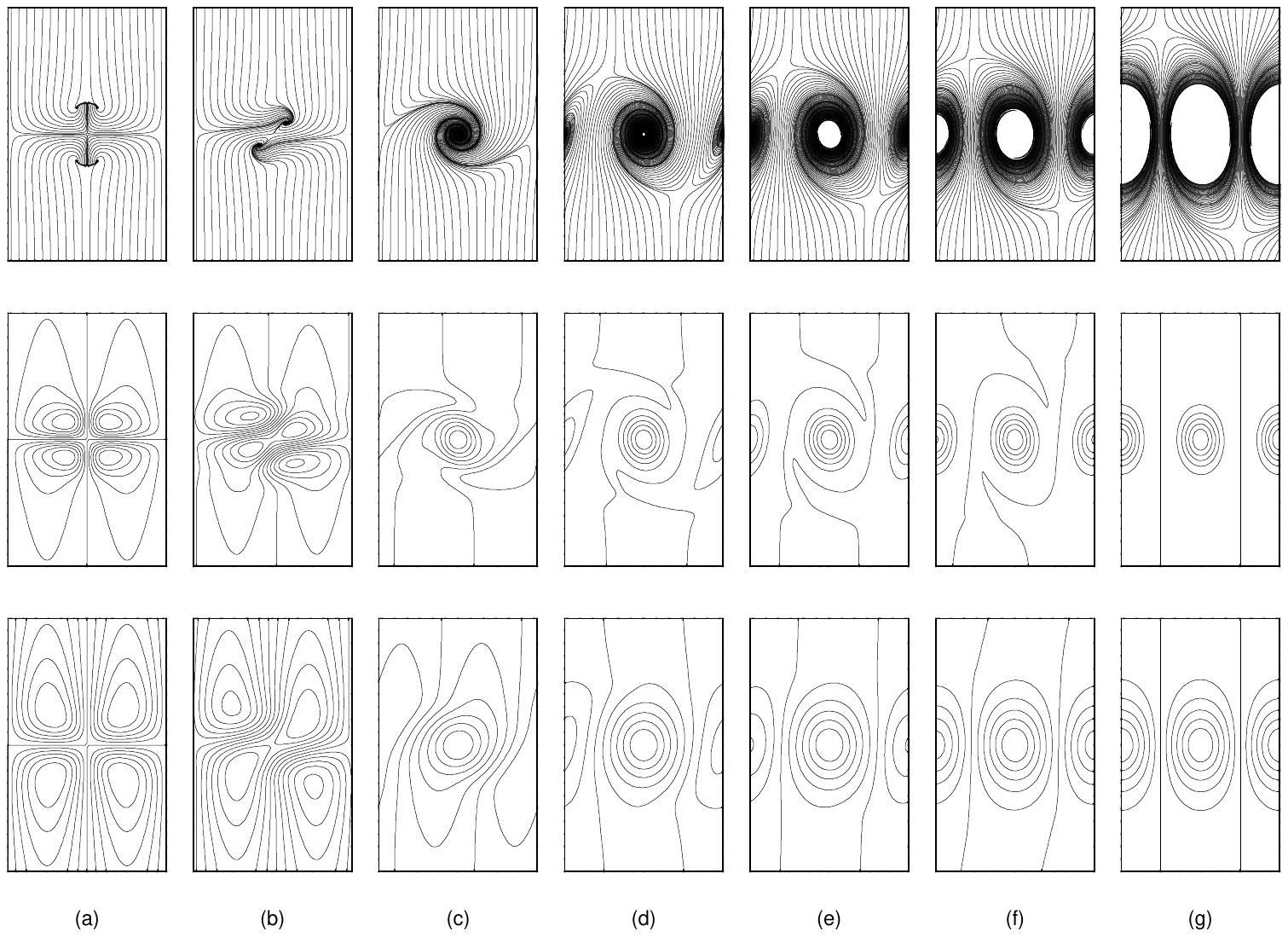} }
\caption{Sequence of flows with $A_+=A_-=10$ for a range of
$\Phi$. The top row shows streamlines in the $x$--$z$ plane for the
full flow, the second row $\omega$ and the last $\psi$. In the last two
rows the contour spacing is a sixth on the maximum modulus of $\psi$
and $\omega$ respectively. The values of $\Phi$ and the corresponding
maxima are
({\it a}) $\upi$, 5.90, 16.49,
({\it b}) $0.9\upi$, 6.16, 17.21,
({\it c}) $0.6\upi$, 21.30, 48.69,
({\it d}) $-0.075\upi$, 66.84, 152.0,
({\it e}) $0.25\upi$, 113.6, 251.9,
({\it f}) $0.595\upi$, 175.8, 403.5,
({\it g}) 0, 562.6, 1368.}
\label{Fig5}
\end{figure}

\begin{figure}
\centerline{\includegraphics[trim=10mm 10mm 10mm 130mm,clip,width=3.0in]{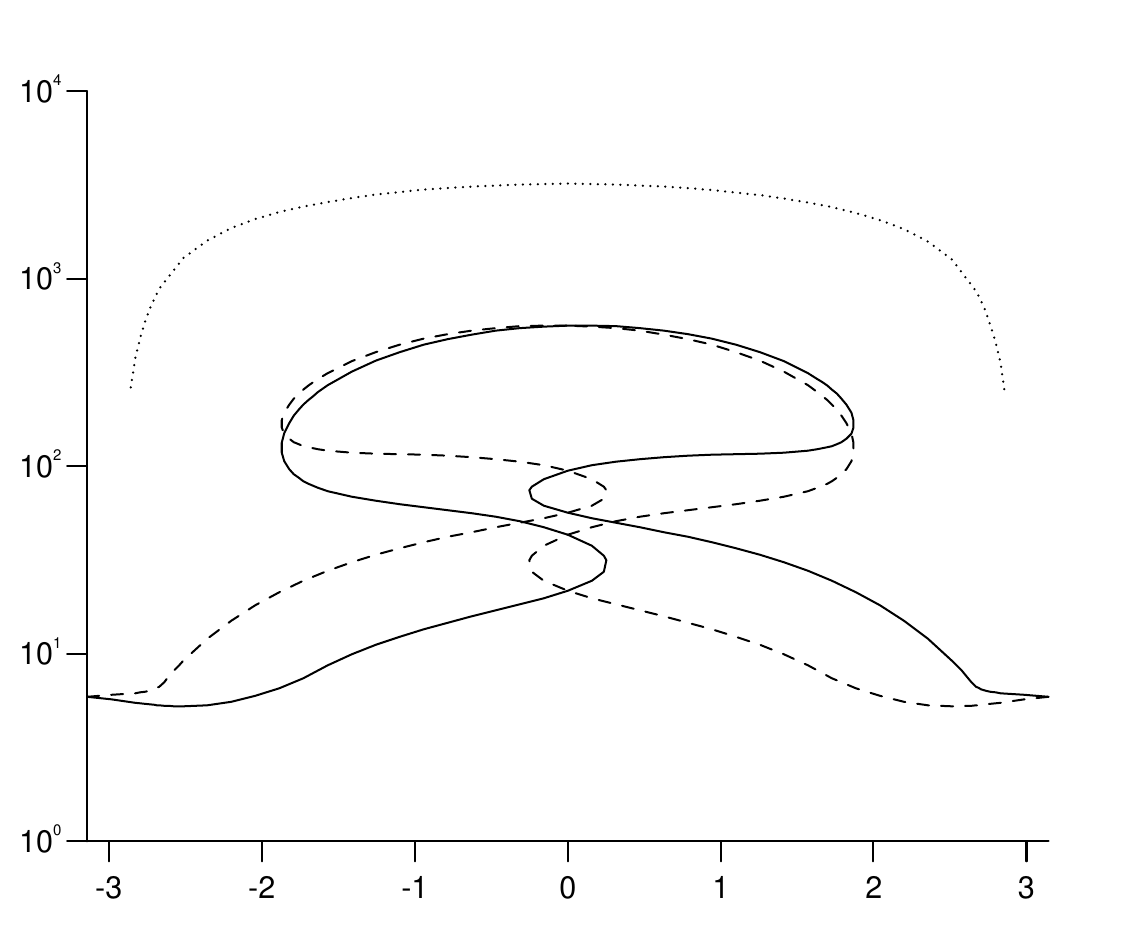} }
\caption{Graph of the sizes of magnitudes of the maxima (solid line)
and minima (dashed line) of $\psi$  for $\lambda=6$, $A_+=A_-=10$ and
$\Phi$ ranging between $-\upi$ and $\upi$. The dotted line shows the
uppermost part of the corresponding curves for $\lambda=8$.}
\label{Fig6}
\end{figure}

\begin{figure}
\centerline{{\includegraphics[trim=10mm 10mm 50mm 130mm,width=3.0in]{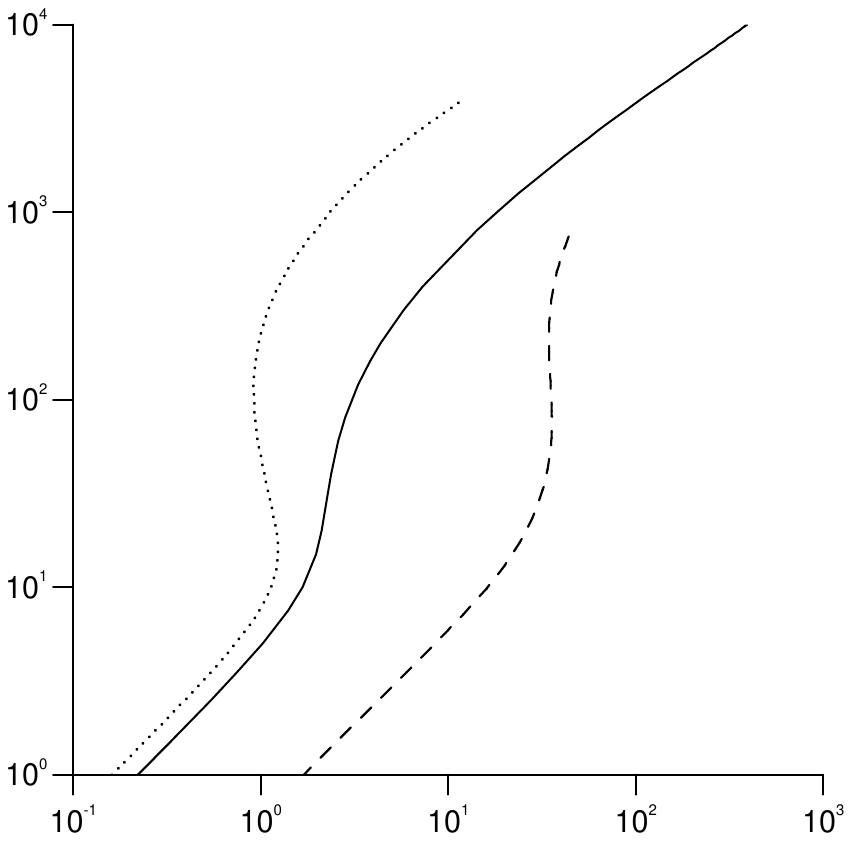}} }
\caption{Graphs of the maximum values of $\psi$ for
vortices  as a function of the tail amplitude $A_+$. The solid and
dotted curves are for KD vortices ($A_+=A_-$, $\Phi=0$) with
$\lambda=6$ and $\lambda=8$ respectively, while the dashed curve is for
the vortices described in section 4.1 ($A_+=-A_-$, $\Phi=0$) with
$\lambda=6$.}
\label{Fig7}
\end{figure}

\begin{figure}
\centerline{{\includegraphics[trim=30mm 40mm 30mm 40mm,width=2.4in]{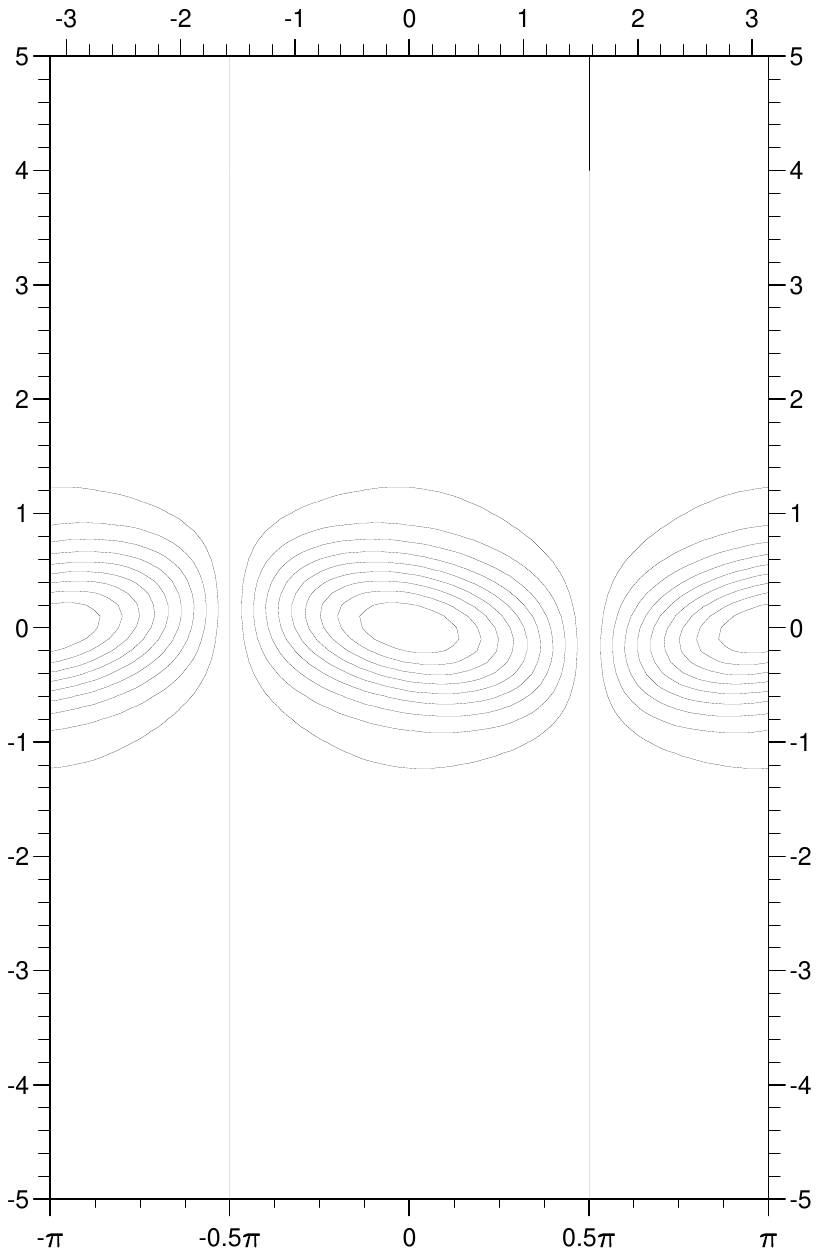}}\quad 
\includegraphics[trim=30mm 40mm 30mm 40mm,width=2.4in]{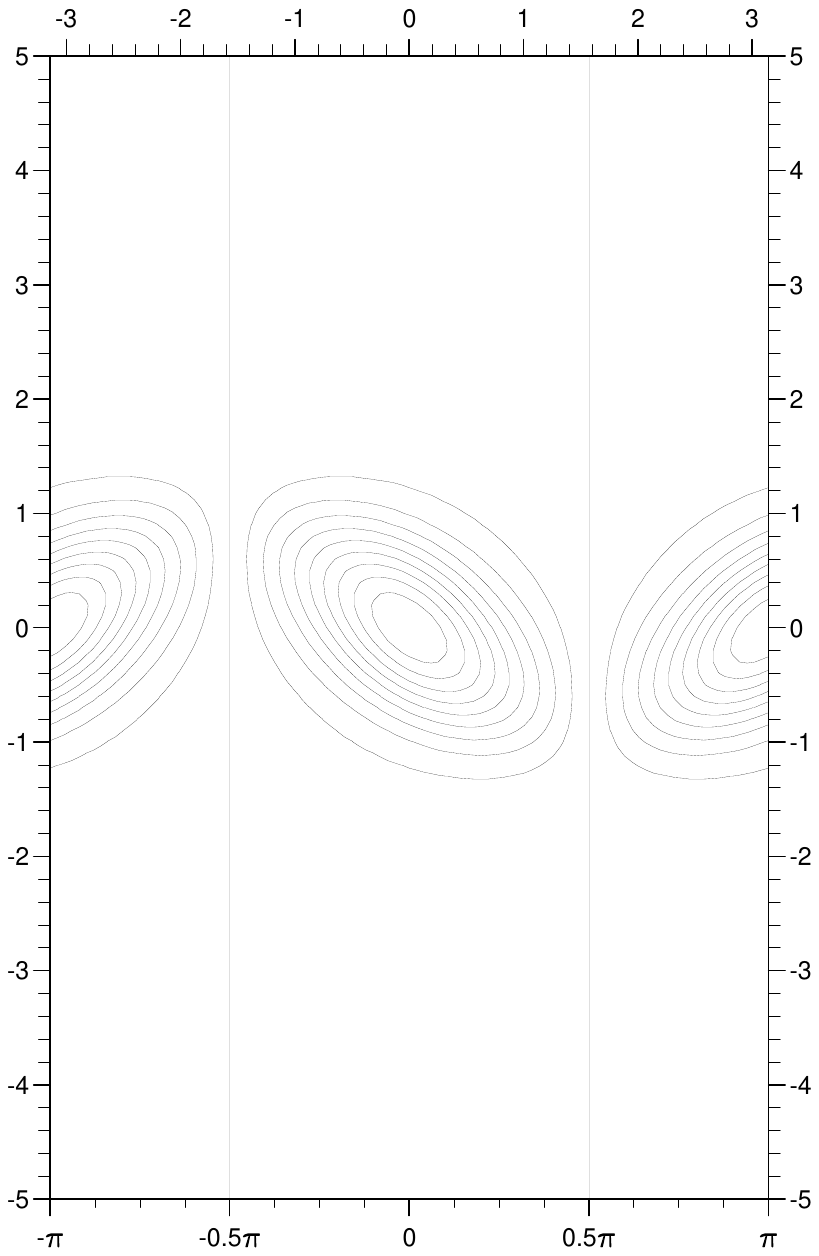}}
\centerline{(a)\hspace{2.4in}(b)}
\centerline{\includegraphics[trim=30mm 40mm 30mm 40mm,width=2.4in]{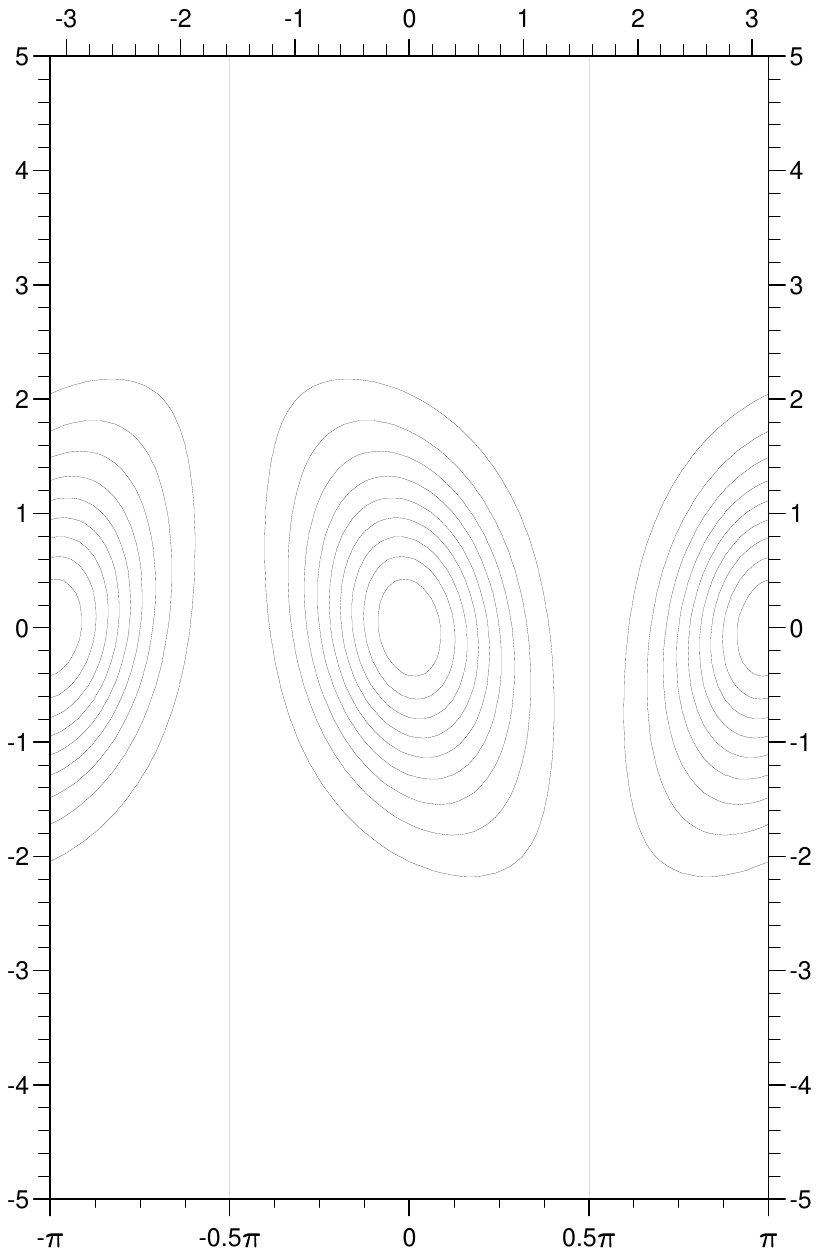}\quad
\includegraphics[trim=30mm 40mm 30mm 40mm,width=2.4in]{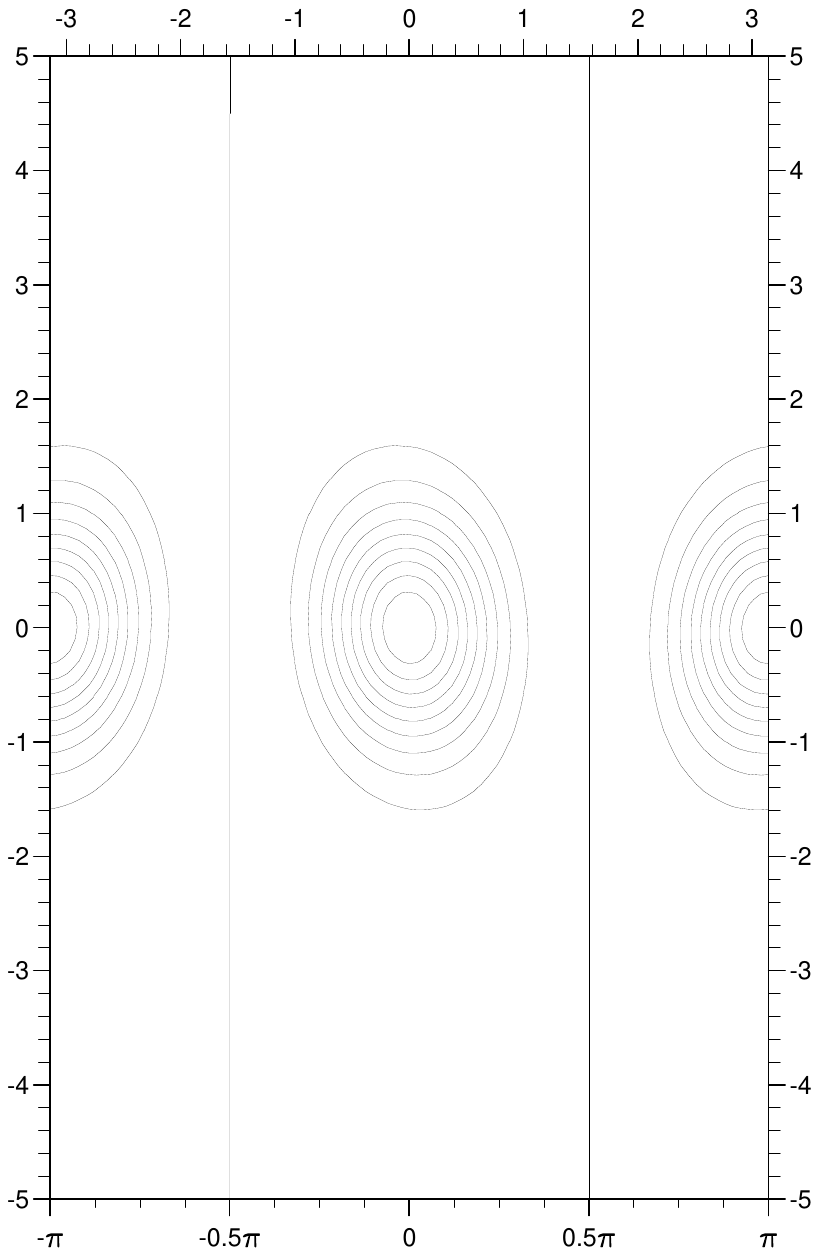}}
\centerline{(c)\hspace{2.4in}(d)}
\caption{Vorticity plots for the vortices corresponding to
figures 5 and 6 of KD. These vortices have amplitudes (a) 2.5, (b) 10,
(c) 40 and (d) 160. The contour spacing is a tenth of the maximum value
of $\omega$. In each case this maximum is (a) 5.696, (b) 23.16, (c)
75.13, and  (d) 388.8.}
\label{Fig8}
\end{figure}

\begin{figure}
\centerline{\includegraphics[trim=30mm 40mm 30mm 40mm,width=2.4in]{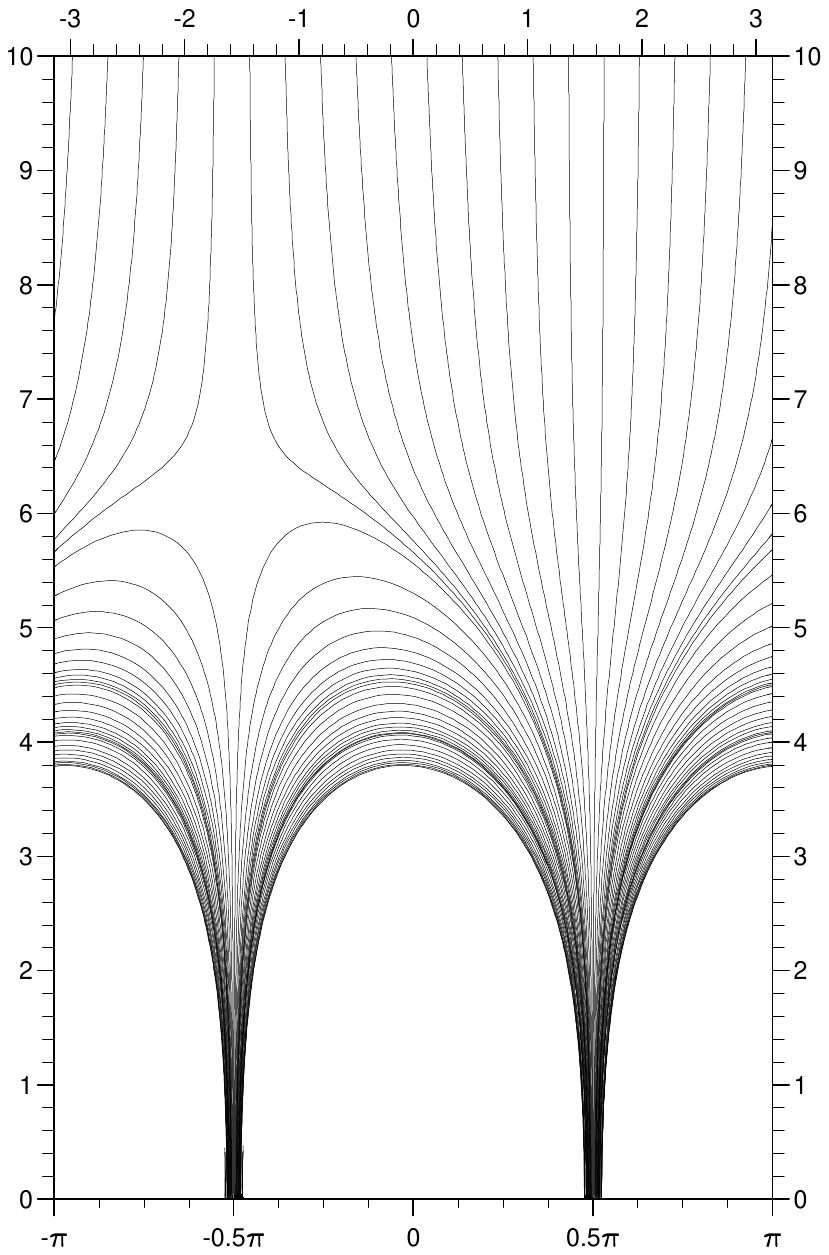}\quad
\includegraphics[trim=30mm 40mm 30mm 40mm,width=2.4in]{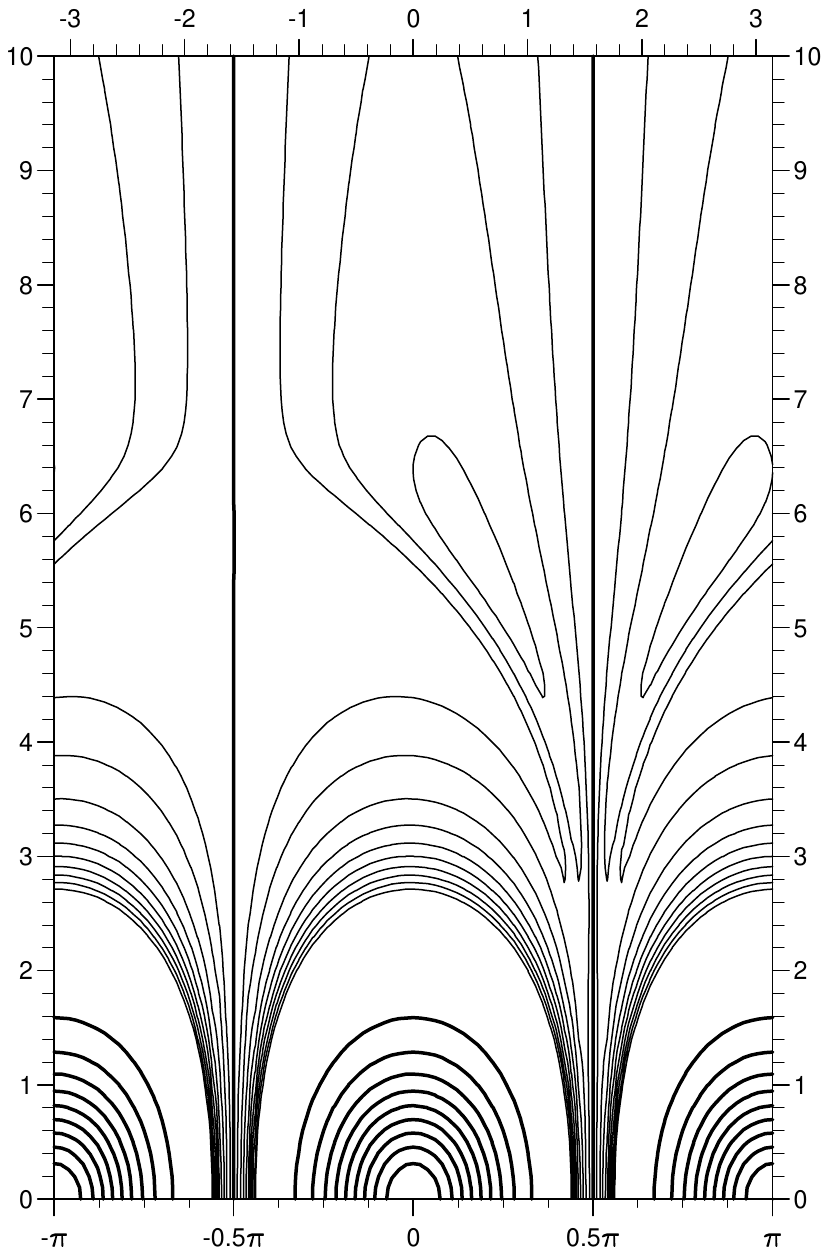}}
\centerline{(a)\hspace{2.4in}(b)}
\caption{Plots of the (a) streamlines and (b) vorticity of
the upper half of vortices with $\lambda=6$ and amplitude 5000. The
contour intervals in (b) are 1216 near the vortex cores, and 10
elsewhere.}
\label{Fig9}
\end{figure}

\begin{figure}
\centerline{\includegraphics[trim=30mm 40mm 30mm 40mm,width=3.0in]{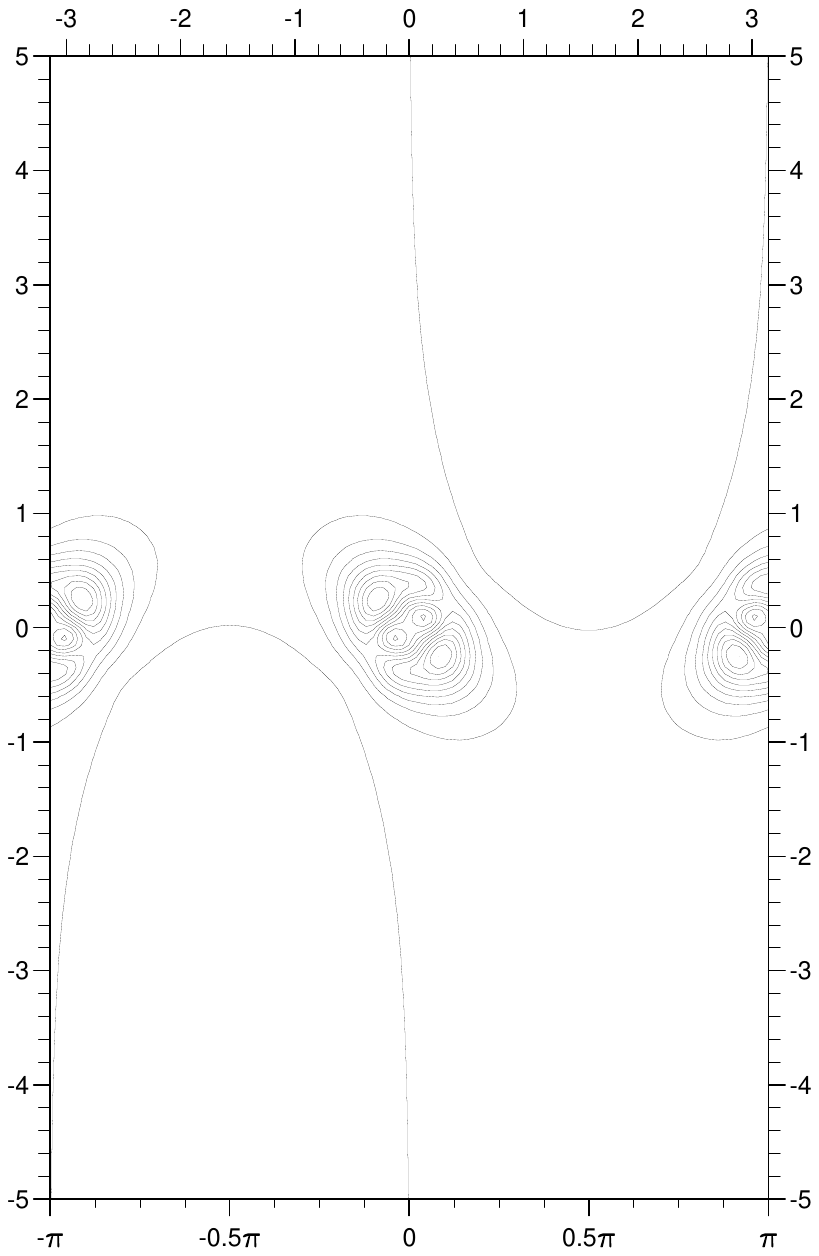} }
\caption{Dissipation for KD vortex with $\lambda=50$ and
amplitude 80. } 
\label{Fig10}
\end{figure}

\end{document}